\documentclass[prl,a4paper,superscriptaddress,twocolumn,showpacs,amsmath,amssymb,floatfix]{revtex4}
\usepackage{amsfonts}
\usepackage{graphicx}

\def\bfp{\mathbf{p}}
\def\bfq{\mathbf{q}}
\def\bfl{\mbox{\boldmath$\lambda$}}
\def\bph{\mbox{\boldmath$\phi$}}

\def\bfxi{\mbox{\boldmath$\xi$}}

\begin{document}
\title{Random matrix ensembles associated with Lax matrices}
\author{E. Bogomolny}
\affiliation{Universit\'e Paris-Sud, LPTMS, UMR8626, B\^at. 100, Universit\'e Paris-Sud, 91405 Orsay, France\\
CNRS,  LPTMS, UMR8626, B\^at. 100, Universit\'e Paris-Sud, 91405 Orsay, France}
\author{O. Giraud}
\affiliation{Universit\'e de Toulouse, UPS; Laboratoire de Physique Th\'eorique (IRSAMC); F-31062 Toulouse, France\\
CNRS; LPT (IRSAMC); F-31062 Toulouse, France}
\author{ \fbox{C. Schmit}}
\noaffiliation
\pacs{05.45.-a, 02.30.Ik}
\date{\today}

\begin{abstract}
A method to generate new classes of random matrix ensembles is proposed. Random matrices from these
ensembles are Lax matrices of classically integrable systems with a certain 
distribution of momenta and coordinates. The existence of an integrable
structure permits to calculate the joint distribution of eigenvalues for
these matrices analytically. Spectral statistics of these ensembles are
quite unusual and in many cases give rigorously new examples of intermediate statistics. 
\end{abstract}

\maketitle

\textit{Introduction:} Statistical properties of surprisingly many different
problems can be  described  by two main distributions: the Poisson
statistics of independent random variables and the Random Matrix Theory
(RMT) statistics. Classically integrable systems display spectral statistics
which are close to Poisson distribution \cite{berry} but classically
chaotic systems, in general, are well described by RMT \cite{bohigas}.
The same classes of spectral statistics appear in the investigation of
disordered systems, in particular in the study of the 3-dimensional
Anderson model (see e.g. \cite{mirlin} and references therein). 
Though the universal character of the Poisson and RMT statistics is well
established (but not always well understood) these two distributions
 do not exhaust all possible behaviors, even in the 3-d Anderson
model. A new phenomenon appears in this model when the disorder strength
is set at a special value corresponding to the metal-insulator transition. 
In \cite{shklovskii} it was established numerically that in this case
spectral statistics has features intermediate between Poisson and RMT
behaviors.  Similar hybrid statistics has  been observed numerically
\cite {Gerland} for certain dynamical systems which are neither integrable
nor chaotic but belong to the class of 
pseudo-integrable systems \cite{berryrichens}.   

These and other examples demonstrate the possible existence of a not yet 
well-defined class of intermediate spectral statistics \cite{mirlin}
with two
characteristic  features: level repulsion at small spacings as in RMT
and  exponential decrease of the nearest-neighbor distribution at large
spacings as in the Poisson distribution. The usual random matrix
ensembles are chosen in such  a way that their measure is invariant under
conjugation $\mathcal{M}\longrightarrow U\,\mathcal{M}\,U^{-1}$
over a group  of either unitary, orthogonal, or symplectic matrices \cite{mehta}.   
The invariance of eigenvalues of $\mathcal{M}$ under these groups permits to find  the exact joint distribution of all eigenvalues $\lambda_j$. In the simplest setting \cite{mehta} 
the distribution reads
\begin{equation}
P(\bfl)\sim \exp\left [-a\sum_{k}\lambda_k^2+\beta\sum_{j<k}\ln |\lambda_j-\lambda_k|\right ]
\label{standard}
\end{equation}
where $\beta=1,2,4$ for, respectively, orthogonal, unitary and symplectic ensembles. 

For intermediate statistics the situation is different. Physical problems
giving rise to intermediate statistics have a natural basis in which they are
defined, and in general do not possess explicit invariant measure, which
makes the  progress of their analytical treatment difficult.

In this letter we introduce new families of random matrix ensembles which
are not invariant over geometrical transformations, but still allow to
obtain an exact joint distribution of eigenvalues analogous to
(\ref{standard}). These ensembles give new non-trivial 
examples of intermediate statistics.\\

\textit{General construction:} To define our random matrix ensembles we consider 
a classical one-dimensional $N$-body integrable model such
that the equations of motion are equivalent to the matrix equation
\begin{equation}
\dot{L}=M\,L-L\,M
\end{equation}
for a pair of Lax matrices $L$ and $M$ depending on momenta $\bfp$ 
and coordinates $\bfq$ \cite{lax}.

It is the Lax matrix $L=L(\bfp\, ,\bfq\, )$ that we propose to
consider as a random matrix depending on random variables $\bfp$ and $\bfq$
distributed according to a certain "natural" measure
\begin{equation}
\mathrm{d}L=P(\bfp\, ,\bfq\, )\, \mathrm{d}\bfp\,\mathrm{d}\bfq\ .
\label{measure_pq}
\end{equation} 
which depends on the system. The only information we shall use from the
integrability of the underlying classical system is the existence and
explicit form of action-angle variables $I_{\alpha}(\bfp\, ,\bfq\, )$ and 
$\phi_{\alpha}(\bfp\, ,\bfq\, )$, and the identity
\begin{equation}
\prod_j\mathrm{d}p_j\,\mathrm{d}q_j=\prod_{\alpha}\mathrm{d}I_{\alpha}\,\mathrm{d}\phi_{\alpha}\ .
\end{equation}
due to the canonicity of the action-angle transformation.
Direct proof of this key identity is difficult and implicit methods were used to establish it \cite{I,III}.
Action variables turn out to be usually the eigenvalues $\lambda_{\alpha}$ of the Lax
matrix or a simple function of them. The canonical change of variables in
(\ref{measure_pq}) from momenta and coordinates to action-angle variables
leads to a formal relation
\begin{equation}
\mathrm{d}L=\cal{P}(\bfl\, ,\bph\, )\, \mathrm{d}\bfl\,\mathrm{d}\bph\ .
\label{measure_lphi}
\end{equation}
The exact joint distribution of eigenvalues is then obtained by integration
over angle variables, which can easily be performed in all cases considered:
\begin{equation}
P(\bfl)=\int \cal{P}(\bfl\, ,\bph\, )\mathrm{d}\bph\ . 
\end{equation} 
This scheme is general and can be adapted to several different models. 
Due to space restrictions we consider here only two representative 
ensembles, based on the rational
Calogero-Moser (CM) \cite{calogero} and the trigonometric
Ruijsenaars-Schneider (RS) \cite{rs} models. Other examples and details of
the calculations will be presented elsewhere \cite{papierlong}.\\

\textit{Calogero-Moser ensemble:} 
The Hamiltonian of the rational CM model reads
\begin{equation}
H(\bfp\, ,\bfq\,)=\frac{1}{2}\sum_k p_k^2+g^2\sum_{i<j}\frac{1}{(q_j-q_i)^2}
\end{equation}
with the following $N\times N$ Hermitian Lax matrix \cite{perelomov} 
\begin{equation}
L_{jk}=p_j\delta_{jk}+\mathrm{i}g \frac{1-\delta_{jk}}{q_j-q_k}\ .
\label{calogero}
\end{equation}
Let $\lambda_{\alpha}$ and $u_k(\alpha)$ be eigenvalues and right
eigenvectors of $L$.
In \cite{I} it is proved that the matrix $Q$ (called conjugate to $L$),
defined by 
\begin{equation}
Q_{\alpha \beta}=\sum_k u_k^{*}(\alpha)q_k u_k(\beta)\ ,
\label{matrix_A}
\end{equation}
can be written as
\begin{equation}
\label{matrix_Qcs}
Q_{\alpha \beta}=\phi_{\alpha}\delta_{\alpha \beta}-\mathrm{i}g
\frac{1-\delta_{\alpha \beta}}{\lambda_{\alpha}-\lambda_{\beta}},
\end{equation}
where the new variables  $\phi_{\alpha}=Q_{\alpha \alpha}$ are angle
variables canonically conjugated to the action variables $\lambda_{\alpha}$.
Equation \eqref{matrix_Qcs} is similar to \eqref{calogero} with the 
substitution $g\to -g$, $q_j\to$ eigenvalues $\lambda_{\alpha}$ and $p_j\to$  
angle variables $\phi_{\alpha}$.

Consider now an ensemble of random matrices (\ref{calogero}) defined 
by random variables $p_i$ and $q_j$ with measure 
\begin{equation}
\mathrm{d}L\sim  \exp \left [-a \mathrm{Tr} L^2 -b \sum_k
  q_k^2\right ]\mathrm{d}\bfp\, \mathrm{d}\bfq
\label{measure}
\end{equation}
where $a$ and $b$ are positive constants. Using the described canonical change of
variables and taking into account that 
$\sum_k q_k^2=\mathrm{Tr}\,Q^2$ with $Q$ given by (\ref{matrix_A}), the measure in (\ref{measure}) can be rewritten as
\begin{eqnarray}
&&\mathrm{d}L\sim 
\exp \left [-a \mathrm{Tr}\,L^2-b \mathrm{Tr}\,Q^2 \right ]
\mathrm{d}\bfl\mathrm{d}\bph\\
&&\hspace{-0.5cm}= \exp \left [-a \sum_{\alpha} \lambda_{\alpha}^2-b \sum_{\alpha}
  \phi_{\alpha}^2-\sum_{\alpha\neq\beta}\frac{bg^2}{(\lambda_{\alpha}-\lambda_{\beta})^2} \right ]\mathrm{d}\bfl\, \mathrm{d}\bph\ .\nonumber
\end{eqnarray}
Integration over $\bph$ gives a constant and we are left with the
following exact joint distribution of eigenvalues of the Lax matrix $L$ with measure (\ref{measure})
\begin{equation}
P(\bfl )\sim \exp \left [-a \sum_{\alpha}
  \lambda_{\alpha}^2-b g^2\sum_{\alpha \neq \beta}\frac{1}{(\lambda_{\alpha}-\lambda_{\beta})^2}
\right ]\, . 
\label{distribution}
\end{equation}
According to this formula, eigenvalues of the above CM ensemble behave as a
1-d gas of particles with inverse square inter-particle potential. No
long-range interaction proportional to $\ln|\lambda_{i}-\lambda_{j}|$
is present, in contrast with standard random matrix ensembles (\ref{standard}). 

The fast decrease of inter-particle potential in (\ref{distribution}) with
the distance between particles permits to approximate (see e.g.~\cite{Gerland})
the nearest-neighbor distribution of eigenvalues of the Lax matrix (\ref{calogero}) by the
formula
\begin{equation}
P(s)\approx A\mathrm{e}^{-B/s^2-Cs}
\label{surmise}
\end{equation}
where  $B$ is a fitting constant, and constants $A$ and $C$ are determined
from the normalization conditions. This expression is not exact but may be 
considered as an analog of the Wigner surmise in RMT \cite{mehta}. 

The measure (\ref{measure}) for coordinates $\bfq$ corresponds to $N$
particles with repulsion confined in an interval of the order of
$1/\sqrt{b}$. In order to simplify numerical investigation it is therefore
natural to use the ''picket fence'' approximation 
$q_k\sim k$ with integer $k$. We thus replace the matrix $L$ by a simpler matrix
\begin{equation}
\tilde{L}_{jk}=p_k\delta_{jk}+\mathrm{i}g\frac{1-\delta_{jk}}{2(j-k)}.
\label{simple}
\end{equation}
The decrease as $|j-k|^{-1}$ of non-diagonal elements in $\tilde{L}$ is a characteristic
feature of intermediate systems \cite{mirlin}. In numerical calculations we chose $N$ variables $p_k$ as i.i.d.~random
variables with uniform distribution between $-1$ and $1$. 
\begin{figure}[t!]
\begin{center}
\includegraphics[width=.95\linewidth]{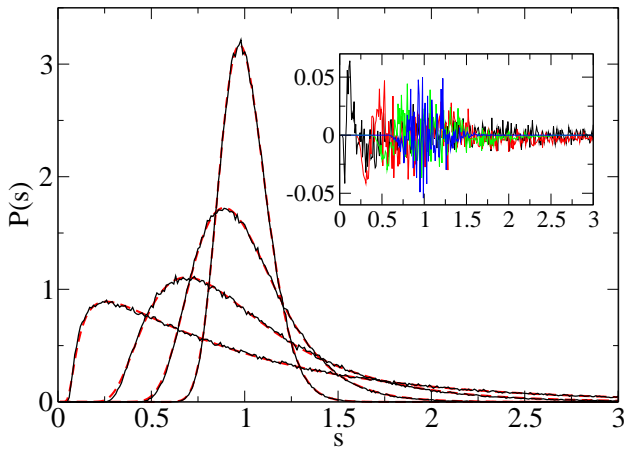}
\end{center}
\caption{Nearest-neighbor distribution for the CM random
  matrices (\ref{simple}) with $g=0.1$, $0.5$, $1$, and $2$ (from left to
  right), averaged over $5000$ realizations of matrices of size $N=301$. 
Black solid lines are numerical results, red dashed lines indicate the fit
  (\ref{surmise}) with fitted values of $B$. For the considered values of
  $g$, $B=0.096$, $0.618$, $1.46$, and $3.11$ respectively. Inset:
  Difference between numerical result and fit for $g=0.1$ (black), $0.5$
  (red), $1$ (green), and $2$ (blue).}
\label{vpInr}
\end{figure}
In Fig.~\ref{vpInr} we show the nearest-neighbor spacing distribution for
the matrix $\tilde{L}$ for several values of the parameter $g$. 
The simple surmise (\ref{surmise}) is practically
indistinguishable from numerical results (see inset), which confirms the
existence of unusual exponentially strong level repulsion in this model.\\

\textit{Ruijsenaars-Schneider ensemble:} 
The second example we  consider here is the trigonometric RS model
\cite{rs}. It appears that the random matrix ensemble that was proposed 
in \cite{GirMarKee04} as a quantization of a pseudo-integrable
interval-exchange map and investigated in \cite{BogSch04}-\cite{Remy}
is a particular case of this model. 

The RS model is determined  by the Hamiltonian 
\begin{equation}
H(\bfp,\bfq)=\sum_{j=1}^N\cos(p_j)V_j^{1/2}(\tau,\bfq)
\label{hamiltonian_rs}
\end{equation}
where $V_j(\tau,\bfq)$ depends on $\bfq$ and a real parameter $\tau$ as
\begin{equation}
V_j(\tau, \bfq)=\prod_{k\neq j}\left (1-\frac{\sin^2\tau}{\sin^2 [(q_j-q_k)/2]}\right )\ .
\label{restrictions}
\end{equation}
The Lax matrix $L(\bfp,\bfq)$ for this model is a unitary matrix given by
\begin{equation}
L_{jk}(\bfp,\bfq)=\mathrm{e}^{\mathrm{i}\tau (N-1)+\mathrm{i}p_j}C_{jk}(\bfq)
\mathrm{e}^{\mathrm{i}(q_k-q_j)/2}
\label{final_G}
\end{equation}
(we choose a phase factor different from the one in \cite{III} in order to
get Eq.~\eqref{simple_rs} when $q_k=2\pi k/N$).
Here $C(\bfq)$ is the orthogonal matrix 
\begin{equation}
C_{jk}(\bfq)=W_j^{1/2}(\tau,\bfq)\frac{\sin \tau }{\sin [(q_j-q_k)/2+\tau ]} W_k^{1/2}(-\tau,\bfq)\ 
\end{equation}
with $W_j(\tau,\bfq)W_j(-\tau,\bfq)=V_j(\tau,\bfq)$ and
\begin{equation}
W_j(\tau,\bfq)=\prod_{k\neq j}\frac{\sin  [(q_j-q_k)/2+\tau  ]}{\sin [(q_j-q_k)/2]}\ .
\end{equation}
Action-angle variables are obtained similarly as for the CM
model. Here one considers \cite{III} the conjugate matrix $Q$ defined by 
\begin{equation}
Q_{\alpha \beta}=\sum_k u_k^{*}(\alpha)\mathrm{e}^{\mathrm{i}q_k} u_k(\beta)
\label{a_rs}
\end{equation}
where $u_k(\alpha)$ are eigenvectors of the Lax matrix (\ref{final_G}) corresponding to eigenvalues $\lambda_{\alpha}=\mathrm{e}^{\mathrm{i}\theta_{\alpha}}$. 
In \cite{III} it is shown that $Q_{\alpha \beta}$ can be written in the form
(\ref{final_G}) with the following substitutions:  $\tau\to -\tau$, $q_m\to$
action variables $\theta_{\alpha}$ and  $p_k\to $ angle variables
$\phi_{\alpha}$ canonically conjugated to $\theta_{\alpha}$. 

The important difference of this model from e.g. the above CM model is that
the Hamiltonian  (\ref{hamiltonian_rs}) and the Lax matrix (\ref{final_G})
are defined not on the whole $\bfq$-space but only on a subset of it where
all $V_j(a,\bfq)$ in (\ref{restrictions}) are positive (notice the square
roots in these expressions). These restrictions depend only on coordinates
and on $\tau$ (in \cite{III} only the case $0<\tau<\pi/N$ had been considered). 
Let $R(\tau,\bfq)$ be the characteristic function of this subset
\begin{equation}
R(\tau,\bfq)=\left \{ \begin{array}{cc}1 &\mathrm{when}\;V_j(\tau,\bfq)>0\ ,\;j=1,\ldots,N\\
0&\mathrm{otherwise}\end{array}\right . .    
\end{equation}
We choose as a "natural" measure for the RS ensemble the uniform
measure of random variables $\bfp$ and $\bfq$ on the region allowed by the 
above restrictions. This implies that the measure on the RS ensemble is chosen as
\begin{equation}
\mathrm{d}L\sim R(\tau,\bfq)\, \mathrm{d}\bfp\, \mathrm{d}\bfq\ .
\label{measure_rs}
\end{equation} 
We transform this expression to action-angle variables, 
and perform the integration over angle variables. Since $Q$ and $L$ have the 
same form but with $\bfl$ and $\bfq$ interchanged, $\bfl$ is subject to the same restrictions
as $\bfq$. We conclude that the exact joint probability of eigenvalues of the ensemble
of random RS matrices (\ref{final_G}) is
\begin{equation}
P(\bfl)\sim R(\tau,\bfl)\ .
\label{joint_rs}
\end{equation}
As mentioned,  (\ref{final_G}) is a generalization of the model investigated
in \cite{BogSch04} and \cite{Remy}. The simplest non-trivial new case  
corresponds to the choice $\tau=\pi b/N$ with fixed $b$. To find
$R(\tau,\bfl)$ in this case we notice that as in \cite{BogSch04} and
\cite{Remy} the matrix (\ref{final_G}) permits two rank-one deformations with
known eigenvectors and eigenvalues
$N_{jk}^{(\pm)}=L_{jk}\mathrm{e}^{\pm \mathrm{i}(q_j-q_k+2\tau)}$.
Generalizing the discussion in  \cite{BogSch04} and \cite{Remy}, one can
prove \cite{papierlong} that for $N$ large enough, there exist exactly 
$n=[b]$ other eigenvalues at angular distance $2\pi b/N$ from any
eigenvalue $\theta_{\alpha}$ (here $[b]$ denotes the integer part of $b$).

Consider an ordered sequence of eigenphases on the unit circle,
$\theta_1<\theta_2<\ldots <\theta_N$ and denote the nearest differences by
$\xi_k=\theta_{k+1}-\theta_k$. Introducing two functions
\begin{equation}
f(x)=\left \{ \begin{array}{cc}1&\mathrm{when}\; 0<x<b\\0&\mathrm{otherwise}
\end{array}\right.,\,\; g(x)=1-f(x),
\end{equation}
one can show \cite{papierlong} that these restrictions give rise to the
following expression for the joint probability (\ref{joint_rs}) of RS Lax matrix eigenphases inside
an interval of length $\Delta$
\begin{equation}
P(\bfxi)\sim \prod_{j=1}^{N}f(s_j)g(s_j+\xi_{j+n})\delta  (\Delta-\sum_{k=1}^N\xi_k )\ ,
\label{joint_general}
\end{equation} 
where $s_j=\xi_{j}+\ldots+\xi_{j+n-1}$ and $n=[b]$.
This formula means that eigenvalues of RS random matrices (\ref{final_G})
with $\tau=\pi b/N$ behave exactly as a 1-d gas where each particle interacts
with $n=[b]$ nearest-neighbors. In is known that in this case all
correlation functions in the limit of large $N$ can be calculated by the
transfer operator method (see e.g. \cite{Gerland}). 
Here we present  a few results for the $k^{\mathrm{th}}$ nearest-neighbor
distributions, $P(k,s)$ which determine the probability that in the interval
of length $s$ there exist exactly $k-1$ other eigenvalues ($P(1,s)\equiv
P(s)$). The details will be discussed elsewhere \cite{papierlong}. 

When $0<b<1$, $P(k,s)=0$ for $0<s<kb$ and for $s>kb$ $P(k,s)$ is a shifted Poisson distribution 
\begin{equation}
P(k,s)=\frac{\mathrm{e}^{-(s-kb)/(1-b)}}{(k-1)!(1-b)^k}(s-kb)^{k-1}\ .
\label{next_nearest01}
\end{equation}   
For larger $b$ formulas, though explicit, become tedious. For example, for $b=4/3$,  $P(1,s)$ is non-zero only when $0<s<4/3$ and  $P(2,s)$ when $4/3<s<8/3$.  Inside these intervals
\begin{equation}
P(s)=\frac{81}{64}s^2,\;
P(2,s)=(-\frac{3}{2}+\frac{27}{16}s-\frac{81}{512}s^3)\mathrm{e}^{3s/4-1}.
\label{nearest}
\end{equation}
To simplify numerical calculations we use (as in (\ref{simple})) the picket
fence approximation of coordinates  $q_k=2\pi k/N$. At
these values of $\bfq$,  $W_j(\tau,\bfq)=\sin N\tau/(N\sin \tau)$ and the
Lax matrix takes the form 
\begin{equation}
\tilde{L}_{jk}=\mathrm{e}^{\mathrm{i}p_k}
\frac{1-\mathrm{e}^{2 \mathrm{i}\tau N}}{N(1-\mathrm{e}^{2\pi
    \mathrm{i}(j-k)/N+2\mathrm{i}\tau})}
\label{simple_rs}
\end{equation}
with random phases $p_k$ uniformly distributed in $[0,2\pi[$.
 
\begin{figure}[t!]
\begin{center}
\includegraphics[width=.95\linewidth]{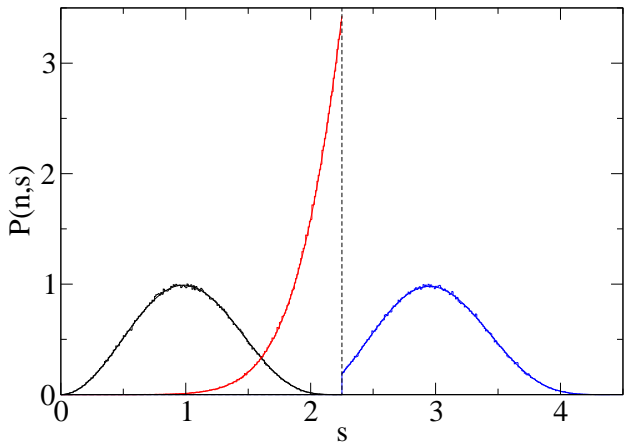}
\end{center}
\caption{Nearest-neighbor distributions (from left to right) $P(1,s)$ (black), $P(2,s)$ (red), and 
$P(3,s)$ (blue) for RS random matrices (\ref{simple_rs}) with $\tau=\pi b/N$ and
  $b=9/4$. All curves correspond to  $N=701$ averaged over $1000$
  realizations of random phases $p_k$. Solid lines are theoretical
  predictions. Dashed vertical line indicates abscissa $9/4$.}
\label{fig9_4}
\end{figure}
When $\tau=\pi \alpha$ with fixed $\alpha$, 
this matrix  up to  notations coincides with the one
proposed in \cite{GirMarKee04} and investigated in \cite{BogSch04} and
\cite{Remy}. We put $\tau=\pi b/N$ and perform numerical calculations of
spectral statistics of matrix (\ref{simple_rs}) for different values of $b$
and find that all above formulas very well agree with numerics. As an
illustration we present at Fig.~\ref{fig9_4} a case with $b=9/4$ for which
explicit formulas are too long to be presented here. Even in this more
complicated case analytical results are difficult to  distinguish from numerics. 

\textit{Conclusion:} To summarize, we proposed a general method of
constructing non-invariant random matrix ensembles whose joint distribution
of eigenvalues can be calculated analytically. These ensembles are Lax
matrices of classically integrable $N$-body models, equipped with a suitably
chosen measure of momenta and coordinates which depends on the model. For such matrix ensembles
 the symmetry groups of
usual RMT are replaced by the underlying structure of integrable flows
generated by $N$ conserved quantities. It is this structure which makes
possible the explicit construction of joint probability of eigenvalues in
these ensembles.  Spectral statistics of these ensembles are quite unusual
and in many cases they present new examples of non-universal intermediate
statistics. In all considered cases eigenfunctions computed numerically
present multifractal properties \cite{papierlong}, which is a typical feature
of intermediate statistics \cite{mirlin}. It is interesting to note that a specific
random matrix ensemble, which appeared in \cite{GirMarKee04} as the result of
quantization of an interval-exchange map, belongs to this class.

\end{document}